\documentclass[11pt, a4paper]{article}
\pdfoutput=1

\usepackage{amsmath}
\usepackage{amssymb}
\usepackage{comment}
\usepackage{multirow}
\usepackage[utf8]{inputenc}
\usepackage{bm}
\usepackage{cite}
\usepackage{physics}
\usepackage{slashed}
\usepackage{braket}
\usepackage{enumitem}
\usepackage{xargs}
\usepackage{tcolorbox}
\usepackage{xfrac}
\usepackage[footnotesize]{caption}
\usepackage{fourier}
\usepackage{cancel}
\usepackage{soul}

\usepackage{ifpdf}
\ifpdf
\usepackage{graphicx, hyperref, xcolor}           
\else                                             
\usepackage[dvipdfmx]{graphicx, hyperref, xcolor} 
\fi
\usepackage{subcaption}

\usepackage{tikz}
\usepackage{tikz-feynman}
\tikzfeynmanset{compat=1.0.0}

\definecolor{rossoferrari}{HTML}{D9073D}
\definecolor{mediumblue}{HTML}{0000CD}
\definecolor{forestgreen}{HTML}{228B22}
\definecolor{desy_blue}{HTML}{009EE2}
\definecolor{desy_orange}{HTML}{FD8800}
\definecolor{light_pink}{rgb}{1,0.4,0.4}
\definecolor{light_blue}{rgb}{0.284602,0.317763,0.963947}
\hypersetup{
setpagesize=false,
bookmarksnumbered=true,%
bookmarksopen=true,%
colorlinks=true,%
linkcolor=light_blue,
urlcolor=mediumblue,
citecolor=forestgreen,
linktocpage=false,
}

\usepackage[height=8.85in,width=6.8in]{geometry}

\makeatletter
\@addtoreset{equation}{section}

\makeatother

\begin{document}
\vspace*{0cm}

\begin{flushright}  CERN-TH-2023-099  \end{flushright}

\vspace*{2cm}

\begin{center}

\vskip 0.0in

{\large \bf Electromagnetic high-frequency gravitational wave detection}

\vskip .3in

Valerie Domcke

\small{ \emph{CERN, Department of Theoretical Physics, 1211 Geneva, Switzerland}}

\end{center}

\begin{abstract}
Ultra-high frequency gravitational waves in the MHz to THz regime promise a unique possibility to probe the very early universe, particle physics at very high energies and exotic astrophysical objects - but achieving the sensitivity required for detection is an immense challenge. This is a brief summary of recent progress in electromagnetic high-frequency gravitational wave searches, which are based on classical electromagnetism in a space-time perturbed by gravitational waves. A particular focus is given to synergies with axion searches and atomic precision measurements. This article was prepared as proceedings for Moriond EW 2023.
\end{abstract}

\vspace*{0.5 cm}

\section{Introduction}

Due to their extremely weak coupling to matter, gravitational waves (GWs) can traverse the universe nearly unperturbed, providing a unique window to probe the epoch before the decoupling of the cosmic microwave background (CMB), when the universe was opaque to electromagnetic (EM) waves. Taking into account cosmic expansion, earlier times and hence higher energies correspond to smaller characteristic physical scales, and cosmological processes at earlier times thus source GWs at higher frequencies, reaching around 100~GHz for thermal processes around the scale of grand unification. At these energy scales, a plethora of different possible extensions of the Standard Model of particle physics are viable and well motivated, some of which entail processes which yield a significant amount of relic GWs.

While probing a stochastic background or GWs from the early universe is a driving motivation for searching for these elusive signals, a more achievable target are exotic astrophysical objects such as mergers of light primordial black holes~\cite{Franciolini:2022htd} or GWs emitted from superradiance of axion clouds around primordial black holes~\cite{Brito:2015oca}. These can lead to locally relatively high energy densities in GWs, facilitating detection if such an object happens to be sufficiently close to the observer. Motivated by this and for simplicity, we will here focus on a toy model comprising of a coherent GW, i.e.\ a monochromatic plane wave.

Identifying the most promising detection strategy for GWs above a kHz remains very much an open challenge, with several concepts summarized in the ultra-high frequency GW living review~\cite{Aggarwal:2020olq}. Here, we will focus on recent progress on EM GW detectors, many of which have been inspired or directly make use of axion searches.

\section{Axion inspired searches}

Starting from electromagnetism with a general metric
\begin{equation}
 S = S_G + S_{EM} =  \int d^4x \sqrt{-g} \left( \frac{1}{2} M_P^2 R - \frac{1}{4} F_{\mu \nu} g^{\alpha \mu} F_{\alpha \beta} g^{\beta \nu} \right)\,,
\end{equation}
with the  field strength tensor $F_{\mu \nu} = \partial_\mu A_\nu - \partial_\nu A_\mu$,
we expand around a flat background perturbed by a gravitational wave, $g_{\mu \nu} = \eta_{\mu \nu} + h_{\mu \nu}$ with $|h_{\mu \nu}|\ll 1$, to obtain the equations of motion of electromagnetism in linearized gravity. Expanding the first term gives~\cite{Maggiore:2007ulw}
\begin{equation}
 S_{G} = - \frac{1}{8} \int d^4 x \left( \partial_\mu h_{\alpha \beta} \partial^\mu h^{\alpha \beta} - (\partial_\mu h) (\partial^\mu h) + 2 \partial_\mu h^{\mu \nu} \partial_\nu h - 2 \partial_\mu h^{\mu \nu} \partial_\rho h^\rho_{\; \nu} \right) + {\cal O}(h^3)\,,
 \label{eq:SG}
\end{equation}
which describes the propagation of GWs with $h \equiv h^\mu_{\; \mu}$. Here and in the following, we have made powers of $h$ explicit, so all indices are understood to be lowered and raised using the flat metric.
The second terms yields~\cite{Herman:2020wao,Berlin:2021txa}
\begin{equation}
 S_{EM} =  \int d^4 x \left(- \frac{1}{4} F_{\mu \nu} F^{\mu \nu} - \frac{1}{2}  j^\mu_{\rm eff} A_\mu \right) + {\cal O}(h^2) \,,
 \label{eq:SEM}
\end{equation}
where the effective current $j_{\rm eff}^\mu$ can be expressed through polarization and magnetization vectors~\cite{Domcke:2022rgu}, 
\begin{equation}
 j^\mu_{\rm eff} = ( - \nabla \cdot {\bf P}, \nabla \times {\bf M} + \partial_t {\bf P})\,,
 \label{eq:jeff}
\end{equation}
with
\begin{equation}
 P_i = - h_{ij} E_j + \frac{1}{2} h E_i + h_{00} E_i - \epsilon_{ijk} h_{0j} B_k\,, \quad M_i = - h_{ij} B_j - \frac{1}{2} h B_i + h_{jj} B_i + \epsilon_{ijk} h_{0j} E_k \,.
 \label{eq:PM}
\end{equation}
This effective current describes the generation of an induced EM field in the presence of a gravitational wave and a background EM field, and thus formally resembles the effective flux generated by axions in the presence of external EM fields.

\paragraph{LC circuit resonators.} Existing and planned low-mass axion haloscopes such as ABRACADABRA~\cite{Kahn:2016aff,Ouellet:2018beu,Salemi:2021gck}, ADMX SLIC~\cite{Crisosto:2019fcj}, BASE~\cite{Devlin:2021fpq}, DMRadio~\cite{DMRadio:2022jfv,DMRadio:2022pkf,DMRadio:2023igr}, SHAFT~\cite{Gramolin:2020ict} and WISPLC~\cite{Zhang:2021bpa} us a resonant LC circuit to search for tiny oscillating induced magnetic fields generated by a dark matter axion background in the presence of a strong external static magnetic field $B_0$. From Eqs.~(\ref{eq:SEM}) - (\ref{eq:PM}) we see that the same setup is in principle also suited for detecting GWs. The resulting magnetic flux induced by a coherent GW can be computed using standard methods of electromagnetism in dielectric media and parametrically scales as~\cite{Domcke:2022rgu,Domcke:2023bat}
\begin{equation}
 \Phi_h \sim e^{- i \omega t} B_0 \,  h  \, (\omega L)^n L^2 \,,
 \label{eq:Phih}
\end{equation}
where $\omega$ denotes the GW frequency, $L$ the characteristic size of the detector and the integer $n \geq 2$ depends on the detector geometry. Since all these experiments operate in the quasi-static limit, $\omega L \ll 1$, smaller values of $n$ imply a better GW sensitivity. However, the cylindrical symmetry often used in axion experiments to maximize the sensitivity to the axion cancels the leading order contribution to the GW induced flux~\cite{Domcke:2023bat}. This can, luckily, be relatively easily remedied by measuring the induced magnetic field with a pick-up loop configuration that breaks the cylindrical symmetry. 

\begin{figure}[t]
\begin{minipage}{1 \textwidth}
\centering
\includegraphics[width=0.8\textwidth]{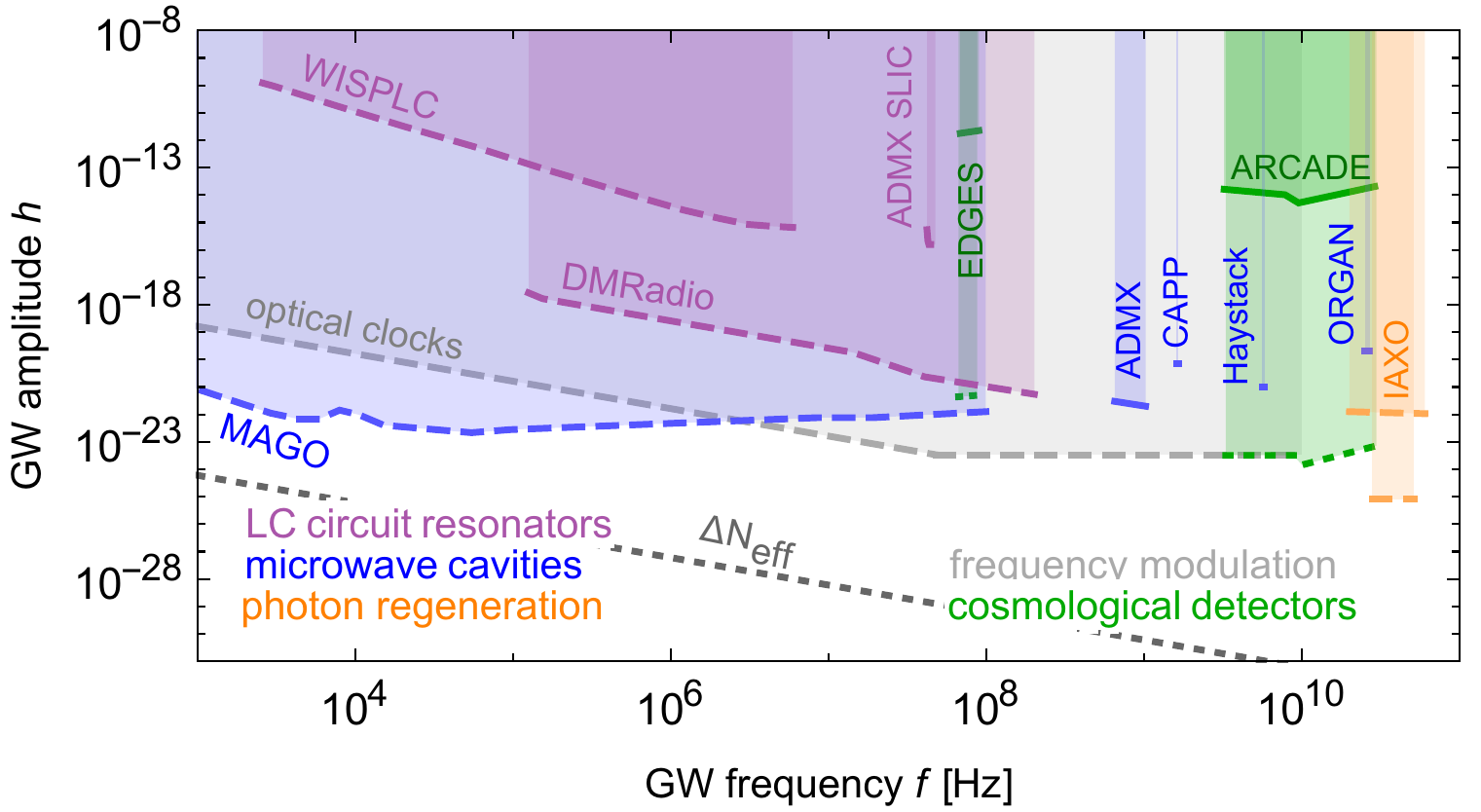}
\caption{Sensitivities of high-frequency EM GW detectors to a coherent GW with amplitude $h$ for existing experiments (solid) and proposed experiments (dashed): microwave
cavities~\cite{Berlin:2021txa,Berlin:2023grv}, LC circuits~\cite{Domcke:2023bat}, frequency modulation~\cite{Bringmann:2023gba}, cosmological
detectors~\cite{Domcke:2020yzq} and photon regeneration experiments~\cite{Ringwald:2020ist}. For reference, the dotted black line shows the cosmological bound on the radiation energy density for an isotropic and stationary GW background.
}
\label{fig:summary}
\end{minipage}
\end{figure}

Bounds on the GW strain from existing and planned axion haloscopes can be obtained by bootstrapping the bounds and predicted sensitivities for the axion searches. The magnetic flux induced by an axion $a$ of mass $m_a$ constituting all of dark matter $\rho_{\rm DM}$ is given by
\begin{equation}
 \Phi_a \sim  e^{- i m_a t} B_0 \, g_{a\gamma\gamma} \, \rho_{\rm DM}^{1/2} L^3 \,.
 \label{eq:Phia}
\end{equation}
Taking into account the differences in duration and coherence, comparing Eqs.~(\ref{eq:Phih}) and (\ref{eq:Phia}) allows a recasting of bounds on $g_{a\gamma\gamma}$ in terms of GW strain $h$~\cite{Domcke:2023bat}. Fig.~\ref{fig:summary} shows (in purple) the recasted limit from ADMX SLIC as well as the expected sensitivity (dashed) of WISPLC and the DMRadio program ($m^3$ and GUT) to a coherent GW~\cite{Domcke:2023bat}.

\paragraph{Microwave cavities.} In axion experiments based on microwave cavities, such as ADMX~\cite{ADMX:2021nhd}, CAPP~\cite{CAPP:2020utb}, HAYSTAC~\cite{HAYSTAC:2018rwy} or ORGAN~\cite{McAllister:2017lkb}, the induced EM field is resonantly enhanced in a cavity with high quality factor $Q_{\rm em}$. The coupling of a GW to an EM resonance  mode yields an induced EM field
$E^{({\rm em})}_h \sim \eta_n Q_{\rm em} \, (\omega L) \, h \, B_0 \, e^{i \omega t}$, where $B_0$ indicates the static externally applied EM field and $\eta_n$ the coupling coefficient between the effective source term~(\ref{eq:jeff}) and the $n$-th cavity mode. Comparing the power spectral density of this signal with the noise power spectral density of the respective instrument gives an estimate of the achievable GW strain sensitivity~\cite{Berlin:2021txa}.

Alternatively, the static external field can be replaced by loading the cavity with a pump mode, as in the MAGO prototype which was designed for GW searches~\cite{Ballantini:2003nt,Ballantini:2005am}. In this case also the mechanical coupling between the GW and the cavity walls is relevant and the induced EM fields are obtained as~\cite{Berlin:2023grv}
\begin{equation}
 E^{({\rm em})}_h \sim Q_{\rm em} \, (\omega L)^2 \, h \, E_0 \, e^{i (\omega_0 \pm \omega) t}  \quad {\rm and} \quad  E^{(\rm mech)}_h \sim Q_{\rm em}\, {\rm min}[1, (\omega L)^2/c_s^2] \, h \, E_0  \, e^{i (\omega_0 \pm \omega) t}\,,
\end{equation}
respectively, where $E_0$ and $\omega_0$ indicate the amplitude and frequency of the externally applied EM field and $c_s$ is the speed of sound of the cavity material. In blue, Fig.~\ref{fig:summary} shows the projected sensitivity of the existing axion experiments ADMX, CAPP, HAYSTAC and ORGAN as well as an estimate of the achievable sensitivity of MAGO~2.0 including the mechanical cavity resonances and assuming that the vibrational noise has been attenuated down the the irreducible thermal component (dashed). A significantly increased sensitivity has been suggested based on reading out the power of the interference term proportional to $E_0 E_h$ instead of $E_h^2$~\cite{Herman:2020wao}, it remains however disputed how this could be concretely achieved~\cite{Berlin:2021txa}.

\paragraph{Photon regeneration experiments.} Photon regeneration experiments such as light-shining-through-the-wall experiments and axion helioscopes rely on the interconversion of axions and EM waves in the presence of an external transverse magnetic field. Similarly, EM waves and GWs undergo oscillations according to the (inverse) Gertsenshtein effect, as can be seen by deriving the equations of motion from Eqs.~(\ref{eq:SG}) and (\ref{eq:SEM}). In particular, the second term in Eq.~(\ref{eq:SEM}) acts as a source term for the EM wave (GW) which is proportional to the GW (EM wave) amplitude. Working for simplicity in the transverse traceless (TT) gauge~\cite{Raffelt:1987im}, 
\begin{equation}
 \left( \partial_t^2 - \partial_\ell^2 + m_\gamma^2 \right) A_\lambda = - B_0 \partial_\ell h^{TT}_\lambda \,, \quad (\partial_t^2 - \partial_\ell^2) h^{TT}_\lambda = \kappa^2 B_0 \partial_\ell A_\lambda \,,
\end{equation}
with $\ell$ indicating the propagation direction, $\kappa^2 = 16 \pi G_N$, $\lambda = \{+, \times \}$ and $m_\gamma$ is the effective photon mass. From this, the conversion probability of GWs to photons after a propagation distance $\Delta \ell$ in the limit $m_\gamma \ll \omega$ is obtained as 
\begin{equation}
 P(\Delta \ell) \simeq \frac{1}{4} \kappa^2 \,  B_0^2 \, \ell_{\rm osc}^{2} \, \sin^{2}\left(\frac{\Delta\ell}{\ell_{\rm osc}}\right) \simeq \frac{1}{4} \kappa^2 \, B_0^2 \, \Delta \ell^2 \,,
\label{eq:GWP1}
\end{equation}
with the oscillation length $\ell_{\rm osc} = 2/\sqrt{m_\gamma^4/(4 \omega^2) + \kappa^2 B_0^2}$. Based on this, limits and projected sensitivities of ALPS~\cite{ALPS:2009des,Bahre:2013ywa}, CAST~\cite{GraciaGarza:2015sos}, IAXO~\cite{Armengaud:2014gea}, JURA~\cite{Beacham:2019nyx} and OSQAR~\cite{OSQAR:2013jqp,OSQAR:2015qdv} have been used to set limits on the energy density of a high-frequency stochastic gravitational wave background~\cite{Ejlli:2019bqj,Ringwald:2020ist}, $\rho_{\rm SGWB} =  4 \pi^2 \int d\ln f \, f^2 h_c^2$. Comparing this to the energy of a coherent GW, $\rho \sim 2 h^2 \omega^2 / \kappa^2$, we see that for a broadband detector, $\Delta f \sim f$, the bounds on $h_c$ can equivalently be taken as bounds on the amplitude $h$ of a coherent signal. Existing axion experiments target optical or x-ray frequencies, however proposals have been made to lower this to the GHz range~\cite{Ringwald:2020ist}, depicted as dashed orange lines in Fig.~\ref{fig:summary}.

\section{Frequency modulation}

An alternative way to use EM waves as a probe of GWs is the impact of GWs on the propagation of the EM wave as well on the position of the EM emitter and absorber. Contrary to the methods described in the previous section, this calculation is linear in the EM field, i.e.\ no background EM field appears. The most commonly known effect is the one exploited in interferometers such as LIGO/VIRGO/KAGRA and LISA, as well as in the Holometer~\cite{Holometer:2016qoh} targeting the high-frequency regime. The GW changes the proper distance in the interferometer arms, resulting in a measurable phase shift (at constant photon frequency). 

Here, we focus on a different effect, which is the modulation of the photon frequency, $\omega_\gamma = - g_{\mu \nu} p^\mu u^\nu$, with $p^\mu$ the photon momentum and $u^\nu$ the four-velocity of the observer. All three quantities on the right hand side can obtain corrections due to a passing GW, with the change in $p^\mu$ described by the geodesic equation and the change in $u^\nu$ depends on the boundary conditions of the sender ($S$) and detector ($D$), i.e.\ if these are freely falling or rigidly mounted. For example, for sender and detector in free fall the relative frequency shift is~\cite{Kolkowitz:2016wyg,Lobato:2021ffr,Bringmann:2023gba}
\begin{equation}
     \frac{\omega_\gamma^D - \omega_\gamma^S}{\omega_\gamma^D}
        = h_+ \cos^2(\vartheta/2)             
        \Big\{\!
                 \cos \varphi^S(t)- \cos\!\big[\omega L (1 \!-\! \cos \vartheta) + \varphi^S(t) \big]
            \Big\} ,
    \label{eq:fshift_freefall}
\end{equation}
where $\vartheta$ is the angle between the photon and GW propagation direction and $\varphi^S(t) \sim \cos(\omega t)$ is the GW phase at the location of the sender at time $t$. The GW thus leads to a frequency modulation with amplitude $h$ of the photon frequency as measured by the detector.

Atomic clock techniques allow extremely precise optical frequency measurements with remarkable progress over the last years, currently reaching resolutions of the order $\Delta f/f \sim 10^{-18}$~\cite{Bothwell:2021fqe}. However, detecting a MHz - GHz frequency modulation is much more challenging. Liberally using Heisenberg's uncertainty relation, precise frequency measurements require long integration times, $\Delta t \sim \Delta f^{-1}$, which will average away the GW signal since $\Delta t \gg f^{-1}$. A possible solution  is an `optical rectifier', i.e.\ a shutter which only allows photons to pass for a fraction of the GW period~\cite{Bringmann:2023gba}. An estimate for the achievable sensitivity is shown as dashed gray line in Fig.~\ref{fig:summary}, for a 1m instrument, one day of measurement time, using a mW optical laser and assuming a frequency resolution to static freqeuncy shifts of $\Delta f/f \sim 10^{-21}$.

\section{Cosmological detectors}

Cosmological high-frequency GW detectors combat the tiny conversion probability of GWs to photons (see Eq.~(\ref{eq:GWP1})) by enhancing the `detector volume' to astrophysical or cosmological scales, at the price of reduced knowledge of the environmental conditions compared to laboratory experiments. For example, exploiting the inverse Gertsenshtein effect in large scale cosmic magnetic fields in the dark ages, gravitational waves in the GHz range can be converted into photons detectable by radio telescopes~\cite{Domcke:2020yzq}, such as EDGES~\cite{Bowman:2018yin} and ARCADE~2~\cite{Fixsen:2009xn}. The resulting bound reflects our poor knowledge of the strength of these cosmic magnetic fields: the green solid line in Fig.~\ref{fig:summary} is based on the lower limit on the magnetic fields from blazer observations, the dashed line assumes the upper limit based on CMB observations.

Other astrophysical environments which have recently been considered in a similar spirit are the depletion of CMB radiation into GWs~\cite{Fujita:2020rdx}, as well as GW to photon conversion in neutron stars and the magnetosphere of planets~\cite{Liu:2023mll,Ito:2023fcr}.

\section{Discussion}

Fig.~\ref{fig:summary} summarizes sensitivities (obtained by recasting or achievable by re-analyzing existing experiments) as well as future prospects. Given the range of different experiments, different search strategies and possible GW signals comparing apples with apples is a challenge and requires knowledge of the detector bandwidth, scanning strategy (when applicable), data analysis procedure, duration and coherence of the GW signal. For simplicity, we have chosen here to quote sensitivities to coherent signals as for many of the proposed experiments, this is the first toy model to consider. However, we note that sourcing a GW signal which stays in-band for hours or days at these high frequencies is extremely non-trivial - though not impossible as the example of superradiance demonstrates. More realistic approaches include the introduction of a coherence ratio factor taking into account the limited coherence and duration of the GW~\cite{Domcke:2023bat}, presenting results in terms of the noise power spectral density~\cite{Berlin:2023grv} and using simplified benchmark signals~\cite{Bringmann:2023gba,Domcke:2023bat}. Moving forward, it will be crucial to address this question in a coherent manner.

Another pertinent question is the coordinate system, or gauge, employed. While of course physical observables are invariant under the choice of gauge, once the gauge is fixed it needs to be applied throughout the computation, describing both the GW and the experimental setup. A rigid experimental setup is more easily described in the proper detector frame, where the coordinates are fixed by an idealized rigid ruler. A free-falling experimental setup (as e.g.\ the mirrors in LIGO but also any system where the GW frequency lies above the lowest resonance mode of the system) 
is, on the other hand, more easily described in the TT frame. 

The list of proposals presented here is of course incomplete. We have focused on proposals deriving directly from the interaction of electromagnetism and gravity. It should however be noted that other mechanical proposals such as bulk acoustic wave devices~\cite{Goryachev:2014yra} and levitated sensors~\cite{Aggarwal:2020umq}, have already advanced to dedicated science runs for GW searches in the MHz regime. Moreover, including charged particles and considering the impact of a passing GW on the Dirac equation has lead to the proposal of magnon detectors~\cite{Ito:2020wxi,Ito:2022rxn}. A more controversial suggestion is the Gaussian beam proposal~\cite{Li:2000du}, which, if open questions around the current noise estimates can be addressed, promises a very competitive sensitivity.

In summary, Fig.~\ref{fig:summary} demonstrates the increase in interest, activity and achievable sensitivity in recent years. It also shows that a lot of the most competitive proposals are small scale experiments (by the nature of the targeted frequency range), using synergies with other precision measurements or particle physics searches. Expected advances in this domain may thus provide chances for spin-offs leading to significant improvements over current proposals.
However, the path to detecting, or setting bounds on realistic GW signals is still far and very challenging. At the moment, even the most optimistic proposals are a few orders of magnitude away from possible astrophysical coherent signals. 
Reaching the sensitivity needed to probe cosmological stochastic backgrounds is even more challenging, however rewarded by a very strong theory motivation: not only are there a range of models which predict such signals but high-frequency GWs are moreover the only conceivable way of probing many of these ideas.


\paragraph{Acknowledgments.} These proceedings are largely based on Refs.~\cite{Domcke:2022rgu,Domcke:2023bat,Bringmann:2023gba} and \cite{Domcke:2020yzq}, and it is a pleasure to thank my fantastic collaborators T.~Bringmann, C.~Garcia-Cely, E.~Fuchs, J.~Kopp, S.M.~Lee and N.~Rodd. Special thanks also to S.~Ellis, S.M.~Lee, A.~Ringwald and N.~Rodd for valuable comments on the draft of these proceedings.

\vspace{-2mm}


\newpage
\small
\bibliographystyle{utphys}
\bibliography{refs}

\end{document}